\documentclass[twocolumn,aps,pra,groupedaddress,superscriptaddress,amsmath,floatfix,amssymb,showpacs,noeprint,english]{revtex4-2}
\usepackage{graphicx}
\usepackage{dcolumn}
\usepackage{bm}
\usepackage[T1]{fontenc}
\usepackage{textcomp}
\usepackage{mathrsfs}
\usepackage{amsmath}
\usepackage{bbm}
\usepackage[normalem]{ulem}
\usepackage{amsbsy}
\usepackage{amstext}
\usepackage{amssymb}
\usepackage{setspace}
\usepackage{esint}
\usepackage{xcolor,colortbl}
\usepackage{float}
\usepackage[colorlinks,citecolor=blue,urlcolor=blue,linkcolor=blue,hypertexnames=true]{hyperref} 
\usepackage{babel}
\usepackage{booktabs}
\usepackage{tikz}

\begin{document}

\title{Memory in quantum processes with indefinite time direction and causal order}

\author{G\"{o}ktu\u{g} Karpat}
\email{goktug.karpat@ieu.edu.tr}
\affiliation{Department of Physics, Faculty of Arts and Sciences, İzmir University of Economics, İzmir, 35330, Turkey}

\author{Bar\i\c{s} \c{C}akmak}
\email{cakmakb@farmingdale.edu}
\affiliation{Department of Physics, Farmingdale State College—SUNY, Farmingdale, NY 11735, USA}
\affiliation{College of Engineering and Natural Sciences, Bahçeşehir University, Beşiktaş, Istanbul 34353, Turkiye}

\date{\today}

\begin{abstract}

We examine the emergence of dynamical memory effects in quantum processes having indefinite time direction and causal order. In particular, we focus on the class of phase-covariant qubit channels, which encompasses some of the most significant paradigmatic open quantum system models. In order to assess the memory in the time evolution of the system, we utilize the trace distance and the entanglement based measures of non-Markovianity. While the indefinite time direction is obtained through the quantum time flip operation that realizes a coherent superposition of forward and backward processes, the indefinite causal order is achieved via the quantum switch map, which implements two quantum processes in a coherent superposition of their two possible orders. Considering various different families of phase-covariant qubit channels, we demonstrate that, when implemented on memoryless quantum processes, both the quantum time flip and the quantum switch operations can generate memory effects in the dynamics according to the trace distance based measure under certain conditions. On the other hand, with respect to the entanglement based measure, we show that neither the quantum time flip nor the quantum switch could induce dynamical memory for any of the considered phase-covariant channels. 

\end{abstract}

\maketitle

\section{Introduction}

In the classical world we inhabit, we naturally perceive time as flowing in a single and definite direction, despite the fact that the underlying fundamental laws of physics are not incompatible with the reversal of time direction. In fact, in macroscopic settings, time is considered to be an essentially asymmetric parameter which only flows in the forward time direction. It has been very recently argued that this is not necessarily the case in the quantum domain, that is, the roles played by the inputs and outputs of quantum processes can be regarded as symmetric in the sense that it is also possible in principle to formulate processes that receive their inputs in the future and give their outputs in the past, as well as vice versa.

Such a symmetry between the inputs and outputs of quantum processes has been previously discussed in literature in different contexts~\cite{Aharonov1964,Aharonov1990,Aharonov2002,Hardy2007,Oeckl2008,Svetlichny2011,Lloyd2011,Genkina2012,Oreshkov2015,Silva2017}. Remarkably, the authors of Ref.~\cite{Chiribella2022} have introduced a mathematical framework to characterize legitimate quantum operations under time reversal, based on the idea of input-output inversion, which relates a bidirectional forward process to the corresponding backward one through a  symmetry transformation. Furthermore, built upon this framework, they have shown that the forward and backward processes can be put in a coherent superposition, giving rise to a quantum process with indefinite time direction, which cannot be described by quantum operations pertaining to a definite time direction. A prototypical instance of such an operation is the quantum time flip~\cite{Chiribella2022}, which can enable processes, that are accessible in both time directions, to act on systems in a coherent superposition of their forward and backward modes. Using photonic setups, it has been experimentally established that the wining probability of a discrimination game under the implementation of quantum time flip exceeds any other strategy having a definite time direction~\cite{Stromberg2022,Guo2022}. In addition, it has also been proved that indefiniteness of time direction in certain quantum channels can result in information theoretic advantages over channels having fixed time direction~\cite{Liu2023}.

The principle of causality states that the events in the present are results of the events that occurred in the past, and concurrently, the present events become causes for the future events. Nevertheless, it has also been recently demonstrated that quantum mechanics allows events to take place with no definite causal ordering. That is, there exist causally inseparable quantum processes that are not compatible with a fixed order between operations~\cite{Chiribella2013,Oreshkov2012,Brukner2014,Araujo2015,Oreshkov2016,Barrett2021}. A well-studied example of such a quantum process is the quantum switch~\cite{Chiribella2013}, which implements two processes in a coherent superposition of their two alternative orders. Quantum switch, just as the quantum time flip, is more than a theoretical concept and has been experimentally realized and investigated in several studies using a variety of different setups~\cite{Procopio2015,Rubino2017,Guo2020,Cao2023,Rubino2022,Goswami2018,Goswami2020}. In addition, it has been shown that the indefiniteness of causal order, generated by the quantum switch, gives rise to many useful applications in various quantum information protocols, for instance, in quantum metrology~\cite{Zhao2020,Chapeau-Blondeau2021}, quantum channel discrimination~\cite{Chiribella2012}, quantum communication complexity~\cite{Guerin2016} and query complexity~\cite{Colnaghi2012,Araujo2014,Renner2022}, noisy transmission of information~\cite{Ebler2018,Procopio2019,Goswami2020a,Caleffi2020,Bhattacharya2021,Chiribella2021a,Chiribella2021b,Sazim2021,Mukhopadhyay2020}, and quantum thermodynamics~\cite{Felce2020,Guha2020,Simonov2022,Liu2022}.

Over the last two decades, characterization and quantification of non-Markovianity, which stems from the onset of dynamical memory effects in quantum processes, has been one of the most active areas of research in open quantum systems theory~\cite{Rivas2014,Breuer2016,Li2018}. Contrary to the well established notion of non-Markovianity for classical processes~\cite{Vacchini2011}, non-Markovianity has been demonstrated to be a multisided phenomenon in quantum domain, whose different aspects could be identified using numerous different approaches~\cite{Shrikant2023}. Due to the potential advantages that memory effects could provide, non-Markovian quantum processes has received significant attention and investigated in various fields of interest, such as, quantum biology~\cite{Thorwart2009,Chin2013}, quantum key distribution~\cite{Vasile2011}, entanglement generation~\cite{Huelga2012}, quantum metrology~\cite{Chin2012}, information processing~\cite{Bylicka2014} and quantum thermodynamics~\cite{Bylicka2016}. Besides, due to recent advances in reservoir engineering techniques, it has become possible to coherently control quantum systems and experimentally test open system models involving dynamical memory effects~\cite{Li2020}.

In this work, our main aim is to comprehend the consequences of the implementation of the quantum time flip and the quantum switch, which respectively define processes with indefinite time direction and causal order, for the onset of dynamical memory effects. Quantifying the degree of memory with the well-established trace distance and entanglement based measures of non-Markovianity, we determine the conditions under which memory arises in certain families of phase-covariant processes, representing the archetypal open system models of qubits, such as anisotropic depolarizing, generalized amplitude damping and eternally CP-indivisible quantum channels. In particular, we concentrate on both CP-divisible and CP-indivisible families of phase-covariant processes, neither of which could display memory effects according to the here considered measures of non-Markovianity and thus are Markovian according to them. We show that memory effects can emerge dynamically in otherwise memoryless phase-covariant processes through the implementation of the quantum time flip or the quantum switch under certain conditions, with respect to the trace based distance measure. In fact, our results demonstrate that the quantum time flip and quantum switch supermaps themselves insert memory in the dynamics rather than activate it. Nonetheless, no such effect can be observed if we choose to quantify memory effects utilizing the entanglement based measure. Among the other results, we demonstrate that some degree of anisotropy in the depolarizing channel is required for the emergence of memory induced by the indefiniteness of the direction of time through the application of the quantum time flip. Additionally, it is also rather remarkable that the indefiniteness of causal order realized by the quantum switch of identical maps can transform a CP-divisible generalized amplitude damping channel into a non-Markovian quantum process, having unbounded degree of dynamical memory effects.

This paper is organized as follows. Sec.~\ref{sec2} introduces the phase-covariant quantum processes describing the dynamics of the open system models considered in our work. In Sec.~\ref{sec3}, we elaborate on characterization and quantification of memory effects in dynamical quantum process, and define the non-Markovianity quantifiers that we consider. Sec.~\ref{sec4} provides an overview of the quantum time flip and the quantum switch superchannels. In Sec.~\ref{sec5}, we present our main results regarding the emergence of dynamical memory in phase-covariant quantum processes having no definite time direction and causal order. Sec.~\ref{sec6} serves as a summary of our central findings.
    
\section{Phase-covariant processes} \label{sec2}

In this section, we describe the type of dynamical processes to be considered in our investigation. In the theory of open quantum systems, assuming that the system is not coupled to its surrounding environment at the initial time, the time-evolution of a physical system is described by a family of time-parameterized linear dynamical maps $\Lambda(t):\mathcal{B}(\mathcal{H})\rightarrow\mathcal{B}(\mathcal{H})$, $t\geq0$, which satisfies the properties of complete positivity and trace preservation (CPTP), with the initial condition $\Lambda(0)=\mathbb{I}$, where $\mathbb{I}$ is the identity operator. Here, $\mathcal{H}$ denotes a finite dimensional Hilbert space and $\mathcal{B}(\mathcal{H})$ the set of linear operators acting on it. At any later time, the state of the system represented by its density operator $\varrho\in\mathcal{B}(\mathcal{H})$ will be given by 
\begin{equation}
\varrho(t)=\Lambda(t)[\varrho(0)]= \text{tr}_{\text{env}}[U(t)(\varrho(0)\otimes\zeta(0))U^\dagger(t)],
\end{equation}
where $\text{tr}_{\text{env}}$ is the partial trace over the environmental degrees of freedom, $\zeta(0)\in\mathcal{B}(\mathcal{H}^\text{env})$ is the initial density operator of the environment, and $U(t)\in\mathcal{B}(\mathcal{H}\otimes\mathcal{H}^{\text{env}})$ is a unitary time-evolution operator for the open system and its environment~\cite{BreuerPetruccione,Holevo}. We note that, throughout this work, the terms quantum maps, quantum channels and quantum processes are used interchangeably.

\begin{figure}
\centering
\includegraphics[width=0.44\textwidth]{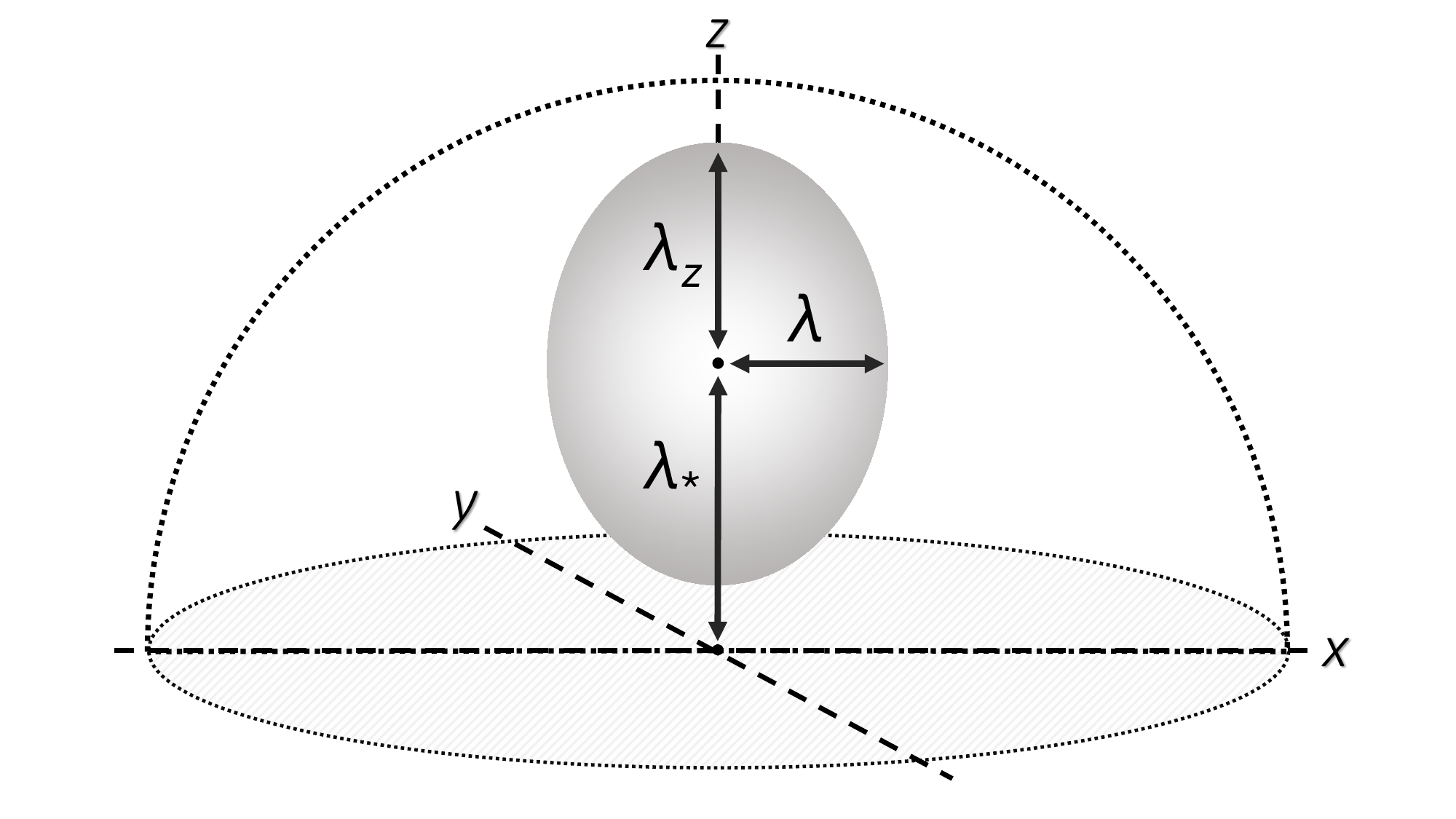}
\caption{Illustrative view of the Bloch sphere under the effect of phase-covariant quantum processes, where we display only the upper half of of the Bloch ball for convenience. Due to the action of the phase-covariant maps, the Bloch ball is deformed, that is, it is contracted in the vertical direction by a factor of $\lambda_z$, in the horizontal plane by a factor of $\lambda$, and finally, its center is displaced by a factor of  $\lambda_*$. Depending on the dynamics of these three real parameters, the ball contracts uniformly, or as a prolate or an oblate spheroid.}
\label{fig1}
\end{figure}

In our analysis, we intend to focus on a class of quantum processes having a particular kind of symmetry, i.e., the covariance property. A quantum process $\Lambda(t)$ is said to be covariant with respect to the unitary representation $V \in\mathcal{B}(\mathcal{H})$ of a group $G$ if
\begin{equation}
\Lambda[V(g) \varrho V^\dagger(g)]=V(g)\Lambda[\varrho]V^\dagger(g),
\end{equation}
for all density operators $\varrho\in\mathcal{B}(\mathcal{H})$ and for all group elements $g$~\cite{Scutaru1979,Holevo1993,Holevo1996}. Conceivably, depending on the unitary representation of the group, covariance property imposes restrictions on the mathematical form of quantum processes. The family of quantum maps that we consider in this work is known as phase-covariant quantum processes for two-level systems (qubits), which are covariant with respect to the phase rotations on the Bloch sphere. Such phase-covariant processes $\Lambda(t):\mathcal{B}(\mathcal{H}_2)\rightarrow\mathcal{B}(\mathcal{H}_2)$ satisfy
\begin{equation}
\Lambda[U(\phi)\varrho U^\dagger(\phi)]=U(\phi)\Lambda[\varrho]U^\dagger(\phi),
\end{equation}
with $U(\phi)=\exp(-i \sigma_z \phi)$ for all $\phi \in \mathbb{R}$ and for any density operator $\varrho \in \mathcal{B}(\mathcal{H}_2)$. As $\mathcal{H}_2$ stands for a 2-dimensional Hilbert space, the standard Pauli operators will henceforth be denoted by $\sigma_x,\sigma_y,\sigma_z \in \mathcal{B}(\mathcal{H}_2)$. Up to an irrelevant unitary rotation, the most general phase-covariant qubit map $\Lambda(t)$ can be expressed as~\cite{Smirne2016,Filippov2020}
\begin{align} \label{pcp}
       \Lambda(t)[\varrho] &= \frac{1}{2}\{\text{tr}[\varrho](\mathbb{I}+\lambda_*(t)\sigma_z)+\lambda(t)\text{tr}[\sigma_x\varrho]\sigma_x \nonumber \\
       &+ \lambda(t)\text{tr}[\sigma_y\varrho]\sigma_y \ + \lambda_z(t)\text{tr}[\sigma_z\varrho]\sigma_z\},
\end{align}
where the three parameters $\lambda(t), \lambda_z(t), \lambda_*(t) \in \mathbb{R}$ satisfy the following eigenvalue equations,
\begin{align} \label{eig}
\begin{gathered}
\Lambda[\sigma_x] = \lambda \sigma_x, \quad \Lambda[\sigma_y] = \lambda \sigma_y,  \quad \Lambda[\sigma_z] = \lambda_z \sigma_z, \\
\Lambda[\varrho_*]=\varrho_*=\frac{1}{2}[\mathbb{I}+\frac{\lambda_*}{1-\lambda_z}\sigma_z ].
\end{gathered}
\end{align}
The time-evolution described by the map in Eq.~(\ref{pcp}) is a valid dynamics, satisfying the conditions required by a CPTP transformation, if and only if~\cite{Filippov2020}
\begin{align} \label{cptpc}
|\lambda_z(t)|+|\lambda_*(t)|\leq1, \hspace{0.25cm} 4\lambda^2(t)+\lambda_*^2(t)\leq [1+\lambda_z(t)]^2.
\end{align}
It can be observed that the last equation in Eq.~(\ref{eig}) dictates the invariant state of the dynamical quantum process, which remains unaffected by the time-evolution. Indeed, it is straightforward to notice that if $\lambda_*=0$, then $\rho_*=\mathbb{I}/2$, that is, we obtain the maximally mixed state, implying that the process is unital. Thus, the degree of unitality of the process is determined by $\lambda_*$. From a geometrical point of view, phase-covariant processes $\Lambda(t)$ transform the Bloch sphere into a spheroid, for which the equatorial radius is $|\lambda|$, the distance from center to pole along the symmetry axis (z-axis) is $|\lambda_z|$, and whose center is shifted by $\lambda_*$ along the z-axis, see Fig.~\ref{fig1}.

It is well-known that a linear map admits an operator-sum (or Kraus) representation if and only if it is CPTP and thus describes a quantum process~\cite{Kraus,Bengtsson}. Then, the action of the phase-covariant channels $\Lambda(t)$ on a density operator $\varrho$ can be obtained as~\cite{Smirne2016}, 
\begin{equation}
\Lambda[\varrho]=\sum_{i=1}^4 M_i \varrho M^\dagger_i,
\end{equation}
where the Kraus operators $\{M_i\}$, satisfying the condition $\sum_i M^\dagger_i M_i=\mathbb{I}$, can be written as
\begin{align}
M_1&=\sqrt{\frac{1-\lambda_z+\lambda_*}{2}}
\begin{pmatrix}
0 & 1 \\
0 & 0 
\end{pmatrix}, \nonumber
M_3=\sqrt{\lambda_+}
\begin{pmatrix}
\cos{\vartheta} & 0 \\
0 &\sin{\vartheta} 
\end{pmatrix}, \nonumber \\
M_2&=\sqrt{\frac{1-\lambda_z-\lambda_*}{2}}
\begin{pmatrix}
0 & 0 \\
1 & 0 
\end{pmatrix}, \nonumber
M_4=\sqrt{\lambda_-}
\begin{pmatrix}
-\sin{\vartheta} & 0 \\
0 &\cos{\vartheta} 
\end{pmatrix},
\end{align}
with the auxiliary parameters $\lambda_\pm$ and $\vartheta$ given by
\begin{align}
\lambda_\pm &= \frac{1+\lambda_z\pm\sqrt{\lambda_*^2+4\lambda^2}}{2}, \nonumber \\
\cot{\vartheta} &= \frac{\lambda_*+\sqrt{\lambda_*^2+4\lambda^2}}{2\lambda}.
\end{align}
It should also be noted that any phase-covariant quantum process $\Lambda(t)$ on qubits can be realized physically in the context of a time-local master equation of the time-dependent Lindblad form~\cite{Lindblad1976},
\begin{align} \label{me}
\frac{d\varrho(t)}{dt} &= \gamma_+(t) \left[\sigma_+\varrho(t)\sigma_- - \frac{1}{2}\{\sigma_- \sigma_+,\varrho(t) \}\right] \nonumber \\
                       &+ \gamma_-(t) \left[\sigma_-\varrho(t)\sigma_+ - \frac{1}{2}\{\sigma_+ \sigma_-,\varrho(t) \}\right] \nonumber \\
                       &+\gamma_z(t)\left[\sigma_z \varrho(t) \sigma_z -\varrho(t)  \right],
\end{align}
where $\sigma_\pm=(1/2)(\sigma_x \pm i\sigma_y)$, and the real-valued time-dependent decoherence rates $\gamma_+(t), \gamma_-(t), \gamma_z(t)$ are linked to the three real-valued parameters, $\lambda(t), \lambda_z(t), \lambda_*(t)$, appearing in Eq.~(\ref{pcp}) in the following way:
\begin{align} \label{dr}
\gamma_+(t) &= \frac{1}{2}\left[ \dot{\lambda}_*(t)-(\dot{\lambda}_z(t)/\lambda_z(t))(\lambda_*+1)     \right], \nonumber \\
\gamma_-(t) &= -\frac{1}{2}\left[ \dot{\lambda}_*(t)+(\dot{\lambda}_z(t))/\lambda_z(t)(1-\lambda_*)  \right], \nonumber \\
\gamma_z(t) &= \frac{1}{4}\left[( \dot{\lambda}_z(t)/\lambda_z(t) - 2\dot{\lambda}(t)/\lambda(t)  \right], 
\end{align}
where $\dot{x}(t)\equiv dx(t)/dt$ denotes the derivative with respect to time. In Eq.~(\ref{me}), as the first and the second terms on the right-hand side describe the energy gain and the energy dissipation processes, respectively, the last term describes pure dephasing. Therefore, it follows that the phase-covariant channels $\Lambda(t)$ embody the combination of these three fundamental physical mechanisms. Actually, some of the most important paradigmatic dynamical quantum processes, such as the amplitude damping ($\gamma_+(t)=\gamma_z(t)=0$), generalized amplitude damping ($\gamma_z(t)=0$), pure dephasing ($\gamma_+(t)=\gamma_-(t)=0$), and depolarizing ($\gamma_+(t)=\gamma_-(t)=2\gamma_z(t)$) channels reside inside the class of phase-covariant quantum maps.

\section{Quantifying memory effects} \label{sec3}

In this section, we will examine the methods that we utilize to characterize dynamical memory effects in quantum processes, describing the dynamics of open quantum systems. There exists a diverse number of techniques in the recent literature to identify the emergence of memory effects in quantum processes~\cite{Shrikant2023}, but these techniques do not always agree on the presence or the degree of memory in the dynamics, since non-Markovianity is known to be a multifaceted phenomenon in quantum theory~\cite{Fanchini2013,Addis2014,Neto2016,Teittinen2018}. Here, we will mainly focus on two different approaches,where two distinct measures of non-Markovianity are constructed, both of which are based on the backflow of information from the environment to the open system, but from different perspectives.

Before we begin to discuss the characterization of memory in dynamical quantum processes, let us first establish what we actually mean by the term memory effects, and its relation to the concept of non-Markovianity. Conventionally, a Markovian quantum process is characterized by a Lindblad type master equation~\cite{Lindblad1976,Gorini1987}
\begin{equation}
\dot{\varrho}(t)=\mathcal{L}\varrho=-i[H,\varrho] +\sum_{i}\gamma_{i}\left[A_i\varrho A_i^\dagger-\frac{1}{2}\left\{A_i^\dagger A_i,\varrho\right\}\right], 
\end{equation}
where the decoherence rates $\gamma_i\geq0$, the noise operators $A_i$ and the Hamiltonian $H$ are all time-independent, generating CPTP maps $\Lambda(t,0)=\exp[\mathcal{L}t]$, $t>0$ that satisfy the semigroup property,
$\Lambda(t_1+t_2,0)=\Lambda(t_1,0)\Lambda(t_2,0)$, for all $t_2,t_1\geq0$. More general kind of quantum maps are described by the time-dependent Lindblad master equation, where $\gamma_i$, $A_i$ and $H$ might explicitly depend on time. Such maps can be written as $\Lambda(t,0)= T \exp[\int_0^t\mathcal{L}(t') dt']$, where $T$ is the time-ordering operator. Provided that all decoherence rates are non-negative throughout the evolution of the system, $\gamma_i(t)\geq0$, these maps satisfy another property, known as CP-divisibility, which states that a CPTP map $\Lambda(t_2,0)$ can be expressed as a concatenation of two other CPTP maps $\Lambda(t_2,t_1)$ and $\Lambda(t_1,0)$ such that
\begin{equation}
\Lambda(t_2,0)=\Lambda(t_2,t_1)\Lambda(t_1,0). \label{cpdiv}
\end{equation}
In the recent literature, quantum processes obeying the CP-divisibility property in Eq.~(\ref{cpdiv}) are conventionally recognized as  Markovian quantum processes. Indeed, $\Lambda(t_2,t_1)$ is a CPTP map if and only if the rates $\gamma_i(t)$ are positive for all $t\geq0$~\cite{Breuer2012}. Hence, CP-divisibility is equivalent to non-negative decay rates in time-dependent Lindblad master equations~\cite{Laine2010}. At this point, it should be emphasized that the decoherence rates $\gamma_i(t)$ can have negative values during the time-evolution and yet the dynamical map $\Lambda(t,0)$ might still be CPTP, describing a legitimate quantum evolution. Nevertheless, in such a situation, the map $\Lambda(t_2,t_1)$ becomes no longer CPTP and thus CP-divisibility property is violated. Such quantum processes, which do not obey the CP-divisibility rule in Eq.~(\ref{cpdiv}), are called CP-indivisible and thus traditionally dubbed as non-Markovian quantum processes. In other words, it is the negativity of decoherence rates $\gamma_i(t)$ and the consequent violation of the CP-divisibility that gives rise to a conventional non-Markovian quantum dynamics.

When it comes to quantification of non-Markovianity, almost all of the measures proposed in literature are actually witnesses rather than strict measures for the violation of the CP-divisibility~\cite{Rivas2014}. To put it in a different way, while they all vanish for CP-divisible dynamics implying a memoryless process, they are not always able to identify the breakdown of the CP-divisibility property. However, some of these non-Markovianity witnesses have still been considered as measures of non-Markovianity or measures of memory effects on their own right. This is due to the fact that they quantify the flow of information from the environment back to the open system, which can make the future states of the open system dependent on its past states, and thus gives rise to memory~\cite{Breuer2009,Fanchini2014,Haseli2014,Karpat2015}. We emphasize that the two quantifiers we introduce here can be regarded as strict measures of non-Markovianity, or equivalently, of memory effects in quantum processes on their own, since both of them have operational interpretations based on information backflow from the surrounding environment to the open system. From this point on, we interchangeably use the terms memory effects and non-Markovianity to indicate the degree of memory effects in quantum processes, as characterized by the quantifiers we specify below.

The first measure we consider in our treatment is constructed upon the distinguishability of a pair of quantum systems that are individually undergoing the same quantum process~\cite{Breuer2009, Laine2010}. This approach interprets the distinguishability between the two initial states of an open quantum system as a manifestation of information flow between the system and its environment based on a protocol introduced in Ref.~~\cite{Gilchrist2005}. When distinguishability monotonically decreases over time for any pair of initial system states, dynamics is characterized as Markovian, signifying a unidirectional information flow from the system to its environment. Conversely, even if a single initial state pair shows a temporary increase in distinguishability during the dynamics, it indicates a backflow of information from the environment to the system, suggesting a non-Markovian evolution. Distinguishability of two quantum systems can be expressed as the trace distance between their density operators $\varrho_1$ and $\varrho_2$ as
\begin{equation} \label{td}
D(\varrho_1,\varrho_2)=\frac{1}{2}\text{tr}\left[(\varrho_1-\varrho_2)^\dag(\varrho_1-\varrho_2)\right]^{1/2},
\end{equation}
which reaches its maximum value for an orthogonal pair of states. Then, based on the trace distance, the degree of memory effects in a quantum process is found as
\begin{equation} \label{tdm}
\mathcal{N}_{D}=\max_{\varrho_1(0),\varrho_2(0)}\int_{\dot{D}(t)>0}\dot{D}(t)dt,
\end{equation}
where the maximization is evaluated for all possible initial state pairs. We note that it has been proved that the optimal state pair for the non-Markovianity measure $\mathcal{N}_{D}$ is always given by a pair of orthogonal states~\cite{Wissmann2012}. Let us here briefly discuss the relation of quantum non-Markovianity based on trace distance and the notion of CP-indivisibility. We first note that all measures of non-Markovianity in the recent literature, including the trace distance measure share a common property, that is, they are all CP-indivisibility witnesses. In other words, by construction they all agree on the fact that CP-divisible processes are memoryless and thus Markovian.  At the same time, we should stress that trace distance measure $\mathcal{N}_{D}$ is not equivalent to the characterization of non-Markovian behavior in quantum processes based on the CP-indivisibility property. In particular, like many other measures of memory effects available in the literature, $\mathcal{N}_{D}$ can vanish and thus indicate the absence of memory in the dynamics for certain CP-indivisible processes due to the lack of information backflow from the environment to the open system. In any case, it is critical to acknowledge that $\mathcal{N}_{D}$ can be employed a measure of memory effects in general, that is, without referring to a time-local master equation of Lindblad form and the notion of CP-divisibility. For instance, the operational usefulness of the measure $\mathcal{N}_{D}$ has been first demonstrated in Ref.~\cite{Laine2014} as a resource for noisy quantum teleportation in a non-local setting, where there exist initial correlations between environmental degrees of freedom~\cite{Liu2013}. Similarly, it could also be reliably used as a measure of memory for open systems described by collision models involving repeated interactions in the strong coupling regime as, for example, discussed in Ref.~\cite{Karpat2021}. Thus,  $\mathcal{N}_{D}$ can be used as meaningful measure of memory on its own right.

Another approach to the characterization of memory in quantum processes is based on the dynamics of quantum entanglement between a principal system of interest and an ancillary system~\cite{Rivas2010}. Let us represent the quantum state of the principal system and the ancilla by the density operators $\varrho_S\in\mathcal{B}(\mathcal{H})$ and $\varrho_A\in\mathcal{B}(\mathcal{H^A})$, respectively, and their composite quantum state by $\varrho_{SA}\in\mathcal{B}(\mathcal{H}\otimes\mathcal{H^A})$. If we suppose that solely the principal system undergoes a process $\Lambda(t_2,0)$ as the ancilla evolves trivially, the CP-divisibility property given in Eq.~(\ref{cpdiv}) and the monotonicity of entanglement under local CPTP maps imply that
\begin{equation}
E[(\Lambda(t_2,0)\otimes \mathbb{I})\varrho_{SA}]\leq E[(\Lambda(t_1,0)\otimes \mathbb{I})\varrho_{SA}]
\end{equation}
for all times $t_2\geq t_1\geq0$, where $E$ is a legitimate entanglement measure. As a consequence, breakdown of the property of CP-divisibility can be identified through the violation of the above inequality. To put it differently, if entanglement between the system and the ancilla shows a temporary revival during the time-evolution, it is understood that the process is not CP-divisible and thus exhibits non-Markovian features. The degree of such memory effects can be quantified by
\begin{equation} \label{ebm}
\mathcal{N}_{E}=\max_{\varrho_{SA}(0)}\int_{\dot{E}_{SA}(t)>0}\dot{E}_{SA}(t)dt,
\end{equation}
where $E_{SA}= E[(\Lambda\otimes\mathbb{I})\varrho_{SA}]$ and the maximization is done over all possible initial states of the composite system. We note in passing that provided that both the principal and the ancillary systems are two-level systems, the optimal state maximizing the measure is given by one of the Bell states~\cite{Neto2016}. Entanglement of formation, which is a monotonic function of concurrence, is defined as $\mathcal{E}=h(1/2+\sqrt{1-C^2}/2)$, with $h(x)=-x\log_2(x)-(1-x)\log_2(1-x)$. Concurrence is given by $C=\max\{0,\lambda_1-\lambda_2-\lambda_3-\lambda_4\}$, where $\lambda_i$ are the square roots of the eigenvalues of $\varrho(\sigma_y\otimes\sigma_y)\varrho^*(\sigma_y\otimes\sigma_y)$ in decreasing order, and $\varrho\in\mathcal{B}(\mathcal{H}_2\otimes\mathcal{H^A_\text{2}})$ in our study. 

Clearly, the entanglement based quantity $\mathcal{N}_{E}$ is a witness of CP-indivisibility, i.e., it measures the degree of CP-divisibility violation, and when it was first introduced in Ref.~\cite{Rivas2010}, it was not in any way related to the flow of information between the open system and its environment. However, later on, it has been demonstrated that, if entanglement is quantified using entanglement of formation~\cite{Wooters2001}, then $\mathcal{N}_{E}$ actually measures the amount of flow of information from the environment back to the open system. In fact, it has been shown that, in a typical tripartite decoherence scenario involving the open system, a measurement apparatus and the environment, the degree of information flow between the open system and its environment can be quantified by the entanglement between the open system and its measurement apparatus. Particularly, through the relation of the entanglement of formation between the open system and the apparatus to the amount of classical information that the environment can access about the open quantum system, known as accessible information, it has been proved both theoretically and experimentally that the emergence of memory in the dynamics due to the back flow of information can be captured by an increase in the entanglement of formation shared by the open system and the apparatus during dynamics of the system~\cite{Fanchini2014,Haseli2014}. For this reason, $\mathcal{N}_{E}$ can be consequently considered as a measure of quantum non-Markovianity on its own right. All in all, even though both the trace distance and entanglement of formation based measures of non-Markovianity, ($\mathcal{N}_{D}$ and $\mathcal{N}_{E}$), are constructed upon the back flow of information, they are simply not equivalent and could always be employed to characterize different aspects of the memory in any legitimate quantum dynamics.

\section{Indefinite time direction and causal order} \label{sec4}
This section is devoted to the description of quantum processes having indefinite time direction and causal order. These quantum channels with no classical analogue are represented by higher order transformations called supermaps or superchannels. Rather than mapping density operators to density operators as ordinary channels do, a superchannel instead takes a quantum channel as an input and outputs another quantum channel.

\subsection{Indefinite Time Direction}

Before introducing the higher order channels with no definite time direction, we briefly discuss the closely related notion of indefinite input-output direction for quantum channels. Let us first note that a quantum process is bidirectional if a forward channel $\Lambda$, which transforms the density operator of the system in a given direction, has an associated valid backward channel $\theta(\Lambda$) mapping the density operator of the system in the opposite direction~\cite{Chiribella2022, Liu2023}. Here the transformation $\theta$ that maps the forward channel into the corresponding backward channel, is an input-output inversion. In fact, a physical process such as the rotation of the polarization of a photon as it traverses through an optical crystal is naturally bidirectional since the photon might pass through the crystal in two opposite directions giving rise to two different quantum processes related by an input-output inversion. All the possible input-output inversions has been characterized in Ref.~\cite{Chiribella2022}, and based on a few physically motivated assumptions, it has been shown that a quantum channel $\Phi$ is bidirectional if and only if it is a doubly stochastic (unital) process with an operator-sum representation 
\begin{equation}
\Phi[\varrho]=\sum_{i} M_i \varrho M^\dagger_i,
\end{equation}
where the requirements $\sum_i M^\dagger_i M_i=\sum_i M_i M^\dagger_i=\mathbb{I}$ are both satisfied. Furthermore, it has also been established that, up to a unitary equivalence, there are only two possible choices for input-output inversion operation, that is, the transpose $\Phi^T[\varrho]=\sum_iM_i^T\varrho\overline{M}_i$ where $\overline{M}_i=(M_i^T)^\dagger$, and the conjugate transpose $\Phi^\dagger[\varrho]=\sum_iM_i^\dagger\varrho M_i$. However, for qubit channels, transpose and conjugate transpose operations are unitarily equivalent implying that the input-output inversion is indeed uniquely defined.

We also note that the input-output inversion is closely related to the conventional time-reversal in quantum mechanics represented by an anti-unitary transformation as argued by Wigner~\cite{Wigner}. Nonetheless, the input-output inversion is actually more general than time-reversal as it can describe processes involving other symmetries. From this point forward, we focus on a particular type of supermap known as the quantum time flip~\cite{Chiribella2022}. This channel takes a bistochastic map $\Phi$ as an input and outputs another bistochastic process $\mathcal{F}_\Phi$, which operates on the density operator of the principal system and an auxiliary qubit that controls the input-output, or equivalently, time direction. The quantum time flip is written as
\begin{equation} \label{qtf}
\mathcal{F}_\Phi(\varrho\otimes\varrho_c)=\sum_i F_i \varrho F_i^\dagger,
\end{equation} 
with the above Kraus operators $F_i$ given by
\begin{equation} \label{qtfk}
F_i=M_i \otimes |0\rangle\langle0|+M_i^T \otimes |1\rangle\langle1|,
\end{equation}
where $\varrho\in\mathcal{B}(\mathcal{H})$ and $\varrho_c\in\mathcal{B}(\mathcal{H^C})$ respectively denote the density operators of the principal system and the control qubit. It is clear that if the control qubit is in the state $|0\rangle$, $\mathcal{F}_\Phi$ acts on the principal system as the forward process $\Phi$. However, if the control qubit starts out in the state $|1\rangle$, then $\mathcal{F}_\Phi$ acts as the backward process $\Phi^T$. The notion of indefinite time direction emerges once the control qubit is initialized in a superposition of the states $|0\rangle$ and $|1\rangle$, in which case $\mathcal{F}_\Phi$ can be seen as a superposition of the forward and the backward channels~\cite{Chiribella2022,Liu2023,Oi2003,Chiribella2019,Abbott2020}. The action of $\mathcal{F}_\Phi$ on the principal system is schematically displayed in Fig.~\ref{fig2}(a). Lastly, we stress that it is impossible to express the quantum time flip channel as a convex mixture of the forward and the backward processes~\cite{Chiribella2022}, thus it indeed describes a coherent superposition of two quantum processes with no definite time direction. 

\subsection{Indefinite Causal Order}

\begin{figure}
\centering
\includegraphics[width=0.41\textwidth]{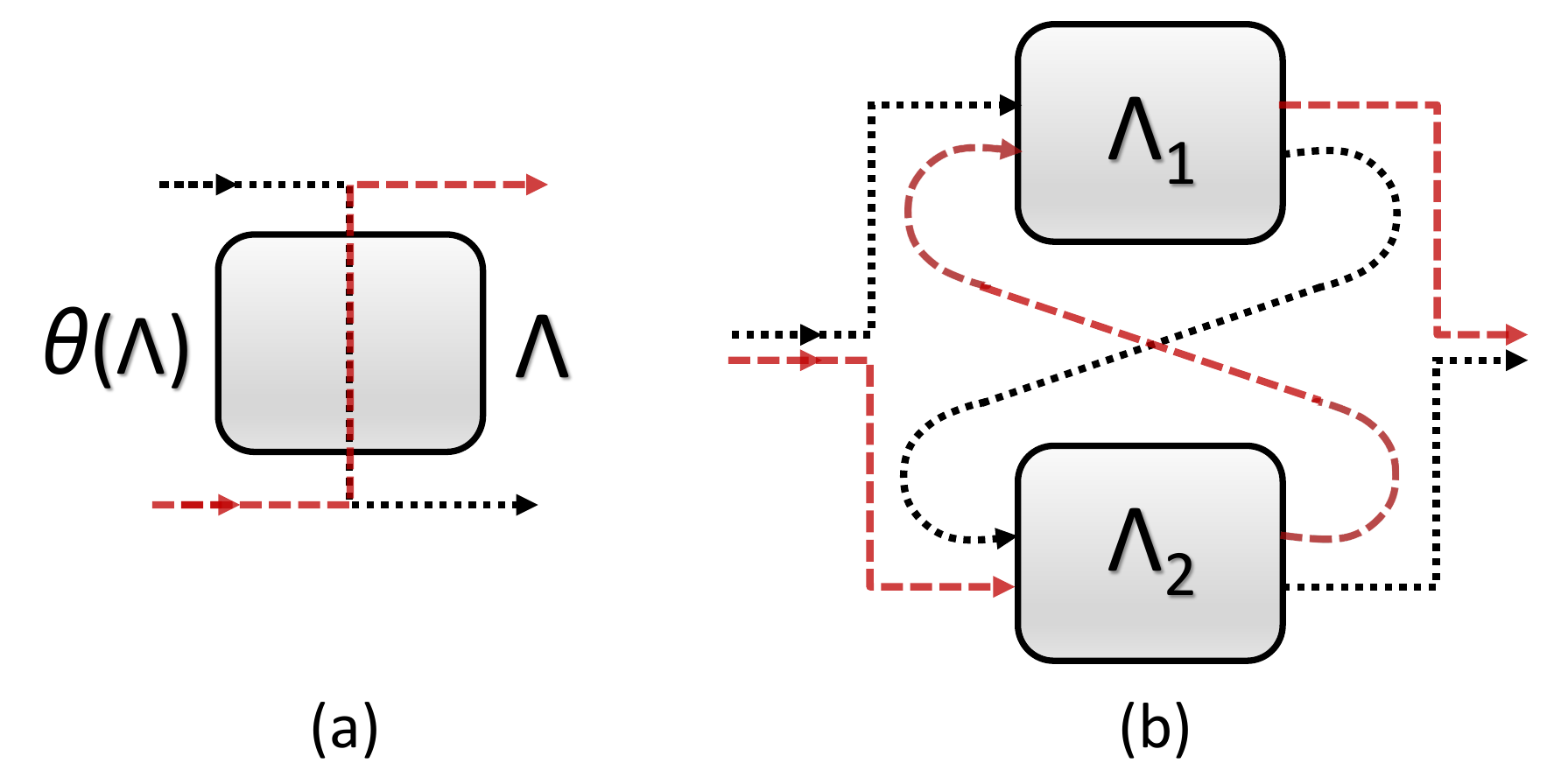}
\caption{(a) Quantum time flip. If the control qubit is initiated in the state $|0\rangle$, the forward channel $\Lambda$ (dotted line) acts on the principal system. Conversely, if the control qubit is in the state $|1\rangle$, the backward channel $\theta(\Lambda)$ (dashed line) affects the system. However, when the control qubit is found in a coherent superposition of the states $|0\rangle$ and $|1\rangle$, then the principal system experiences the quantum time flip, which describes the superposition of the forward and the backward channels, giving rise to a process with indefinite time direction. (b) Quantum switch. If the control qubit starts out in the state $|0\rangle$, the principal qubit is first affected by $\Lambda_1$ and then $\Lambda_2$ (dotted line). Contrarily, when the state of the control qubit is fixed to $|1\rangle$, first $\Lambda_2$ then $\Lambda_1$ acts on the principal qubit (dashed line). Nevertheless, if the control qubit is in a superposition of the states $|0\rangle$ and $|1\rangle$, then the quantum switch implements a superposition of orders  $\Lambda_2\Lambda_1$ and $\Lambda_1\Lambda_2$, resulting in a dynamics with no definite causal order.}
\label{fig2}
\end{figure}

The other superchannel we will consider in our study is known as the quantum switch~\cite{Chiribella2013}, which implements two quantum processes $\Lambda_1$ and $\Lambda_2$ in a coherent superposition of their two possible orders, $\Lambda_1\Lambda_2$ and $\Lambda_2\Lambda_1$. Analogously to the quantum time flip process in Eq.~(\ref{qtf}), the quantum switch makes use of an auxiliary qubit that controls the order of the quantum processes to be implemented on the principal system. If we denote the Kraus operators of $\Lambda_1$ and $\Lambda_2$ by $\{M_i^{(1)}\}$ and $\{M_i^{(2)}\}$, respectively, the Kraus operators of the supermap $S_{ij}$ representing the output of the quantum switch for $\Lambda_1$ and $\Lambda_2$ are given by
\begin{equation} \label{switchkraus}
S_{ij}=M_i^{(2)} M_j^{(1)} \otimes |0\rangle\langle0|+M_j^{(1)} M_i^{(2)} \otimes |1\rangle\langle1|,
\end{equation}
which acts on the composite density operator of the system and the control qubits as
\begin{equation} \label{switch}
\mathcal{S}(\varrho\otimes\varrho_c)=\sum_{i,j}S_{ij}(\varrho\otimes\varrho_c)S_{ij}^\dagger.
\end{equation}
As can be easily observed from the above definition, when the control qubit starts out in the state $|0\rangle$, the supermap $\mathcal{S}$ acts on the principal system implementing the channel $\Lambda_2\Lambda_1$. In contrast, if the initial state of the control qubit is set to $|1\rangle$,  the order of the processes is switched and $\mathcal{S}$ implements the channel $\Lambda_1\Lambda_2$ on the principle system. A particularly compelling quantum effect arises once the control qubit is in a superposition of $|0\rangle$ and $|1\rangle$. That is, the supermap $\mathcal{S}$ given in Eq.~(\ref{switch}) implements a superposition of the two quantum channels $\Lambda_1$ and $\Lambda_2$ simultaneously in two alternative causal orders, or with no definite causal order, as shown in Fig.~\ref{fig2}(b). We should note that the supermap $\mathcal{S}$, which is the output of the quantum switch for quantum processes $\Lambda_1$ and $\Lambda_2$, does not depend on the Kraus representations of the channels $\Lambda_1$ and $\Lambda_2$. Furthermore, its action cannot be replicated by classical mixing or serial application of quantum channels. Hence, the superposition of time-evolutions implemented by the quantum switch generates entirely new dynamics, which cannot be realized considering any other approach pertaining to definite causal order~\cite{Chiribella2013}.

Before moving on to our main results, we would like to elaborate on a crucial point on the nature of the time evolution dictated by the supermaps we consider. The post-selection of the measurement results both in case of quantum time flip and quantum switch in general makes the time-evolution of the density operator $\rho$ non-linear. However, we once again want to highlight the fact that we consider the non-monotonic dynamical behaviors of the trace-distance and entanglement as the main defining property of non-Markovianity or memory effects in quantum processes, based solely on their two distinct interpretations in terms of the information flow between the open system and its environment. Since these two measures only depend on the evolution of quantum states, and not directly on the mathematical properties of the map that describes it, they can be reliably used to characterize the memory behavior of any legitimate CPTP quantum dynamics. Finally, it should be noted that post-selection is one of the main mechanisms which enable the emergence of memory effects after the implementation of quantum time flip and quantum switch. Indeed, as will be seen in the next section, the memory inserted by these two supermaps is due to the combination of two effects. First, the interference between the Kraus operators of the phase-covariant maps as a result of the coherent superposition of evolutions realized by the choice of a coherent initial state for the control qubit, and second, post selection of states after the measurement of the control qubit.

\section{Main Results} \label{sec5}

\subsection{Quantum Time Flip}

We will first focus on the memory effects which can be potentially induced by the quantum time flip in Eq.~(\ref{qtf}), where the bistochastic channel $\Phi$ is described by certain classes of phase-covariant quantum channels. In order to see the quantum effects brought by the superposition of forward and backward processes, we initialize the control qubit in the maximally coherent state $\varrho_c=|+\rangle\langle+|$. After the measurement of the control qubit in the coherent basis defined by the orthogonal projectors $\{|+\rangle\langle+|,|-\rangle\langle-| \}$, where $|\pm\rangle=(|0\rangle\pm|1\rangle)/\sqrt{2}$, we obtain the dynamics of the principal system. Assuming that the outcome corresponding to the measurement operator $|+\rangle\langle+|$ happens, the time-evolution of the principal system reads
\begin{equation}
\varrho(t)=\frac{\text{tr}_c[(\mathbb{I}\otimes|+\rangle\langle+|)\mathcal{F}_\Phi(\varrho\otimes\varrho_c)(\mathbb{I}\otimes|+\rangle\langle+|)]}{\text{tr}[(\mathbb{I}\otimes|+\rangle\langle+|)\mathcal{F}_\Phi(\varrho\otimes\varrho_c)]}.
\end{equation}
In what follows, we consider two different classes of unital phase-covariant processes to analyze the memory behavior of the dynamics induced by the quantum time flip.

\textit{CP-divisible Process}. Let us choose the real parameters defining the phase-covariant process in Eq.~(\ref{pcp}) as 
\begin{equation} \label{dcp}
\lambda(t)=e^{-\omega t}, \quad \lambda_z(t)=e^{- t}, \quad \lambda_*(t)=0,
\end{equation}
where the vanishing of $\lambda_*(t)$ guarantees that the process is unital (doubley stochastic) following the requirement that $\Phi$ is bidirectional. It is straightforward to check that, according to the conditions given in Eq.~(\ref{cptpc}), this process defines a legitimate CPTP quantum evolution provided that $\omega\geq1/2$. In fact, if $\omega=1$, the corresponding channel is nothing other than the isotropic depolarizing channel that describes the uniform contraction of the Bloch sphere down to a single point, i.e., the maximally mixed state $\mathbb{I}/2$ in the origin. Besides, while $1>\omega\geq1/2$ indicates the contraction of the Bloch sphere to the origin as an oblate spheroid, $\omega>1$ corresponds to its contraction as a prolate spheroid, as can be observed in Fig.~\ref{fig1}. On the other hand, the decoherence rates given in Eq.~(\ref{dr}), appearing in the associated Lindblad master equation in Eq.~(\ref{me}), can be calculated as 
\begin{equation}
\gamma_{+}(t)=\gamma_{-}(t)=1/2, \quad \gamma_{z}(t)=(2\omega-1)/4.
\end{equation}
It turns out that none of the decoherence rates are time-dependent and only $\gamma_{z}(t)$ depends on the parameter $\omega$. Since we require that $\omega\geq1/2$ to have a valid dynamics, $\gamma_{z}(t)$ never takes on negative values. Hence, all three decoherence rates are non-negative at all times throughout the evolution of the system, which implies that the process is CP-divisible and thus all of the non-Markovianity quantifiers vanish for this depolarizing channel.

\begin{figure}
\centering
\includegraphics[width=0.375\textwidth]{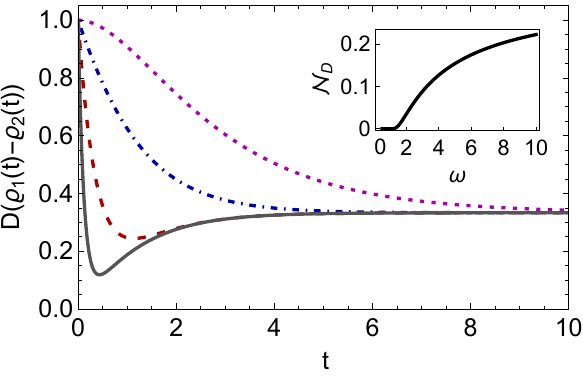}
\caption{Dynamics of the trace distance under the quantum time flip for the CP-divisible process with the initial state pair $\varrho_1=|+\rangle\langle+|$ and $\varrho_2=|-\rangle\langle-|$, supposing that $\omega=0.5$ (dotted line), $\omega=1.0$ (dot-dashed line), $\omega=3.0$ (dashed line) and $\omega=9.0$ (solid line). The inset shows the degree of memory effects quantified via the trace distance measure $\mathcal{N}_D$ as a function of the channel parameter $\omega$.}
\label{fig3}
\end{figure}

Understanding that the considered depolarizing channel cannot exhibit memory effects by itself, we turn our attention to the possibility of the emergence of memory through the implementation of the quantum time flip for this depolarizing process. If the Kraus operators $\{M_i\}$ in Eq.~(\ref{qtfk}) are taken as the Kraus operators of the phase-covariant channel with the parameters given in Eq.~(\ref{dcp}), we obtain the quantum time flip map for the depolarizing channel. At this point, it is important to note that the unital phase-covariant channels we consider are transposition invariant, that is, the time-evolution described by the forward and the backward channels are identical. All the same, the interference between the forward and the backward processes leads to non-trivial outcomes. Let us now consider the trace distance measure and choose the orthogonal initial state pair for the principal system in Eq.~(\ref{td}) as $\varrho_1=|+\rangle\langle+|$ and $\varrho_2=|-\rangle\langle-|$. Then, the dynamics of the trace distance for the depolarizing channel with indefinite time direction is given by
\begin{equation}
D(\varrho_1(t),\varrho_2(t))=\frac{4e^{t(1-\omega)}+e^t-1}{3e^t+1}.
\end{equation}
It is not difficult to show that the above function monotonically decreases in time if $1\geq\omega\geq1/2$. Conversely, when $\omega>1$, there occurs temporary revivals in the dynamics of the trace distance, which signals the emergence of memory effects in the dynamics via the backflow of information from the environment to the principal system. This actually indicates that the quantum time flip supermap inserts the memory in the dynamics rather than activate it as the considered map here is CP-divisible and hence memoryless according to all non-Markovianity criteria. Fig.~\ref{fig3} demonstrates the dynamical behavior of the trace distance for different values of the parameter $\omega$. From a geometrical point of view, the quantum time flip channel gives rise to memory effects only when the bidirectional depolarization process contracts the Bloch sphere to the origin as a prolate spheroid, i.e., when $\omega>1$. Despite we cannot analytically prove that the considered initial state pair is optimal, it appears to be the case, based on the numerical simulations we performed. 

\begin{figure}
\centering
\includegraphics[width=0.375\textwidth]{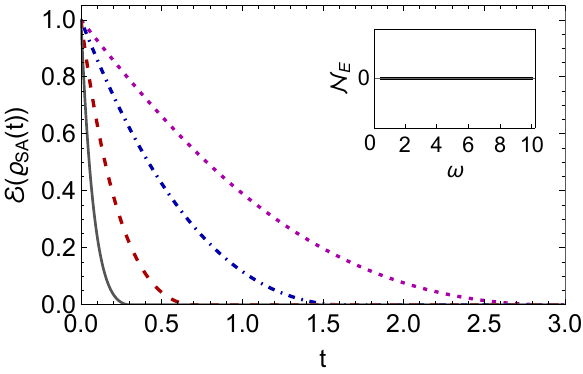}
\caption{Dynamics of the entanglement of formation under the quantum time flip map for the CP-divisible process for the maximally entangled initial state $|\Psi\rangle$, assuming that $\omega=0.5$ (dotted line), $\omega=1.0$ (dot-dashed line), $\omega=3.0$ (dashed line) and $\omega=9.0$ (solid line). The inset shows the degree of memory effects quantified by the entanglement based measure $\mathcal{N}_E$ as a function of the channel parameter $\omega$.}
\label{fig4}
\end{figure}

Having witnessed that the quantum time flip for the depolarization process can induce memory effects, quantified by the trace distance measure, we now consider the entanglement based measure of non-Markovianity. Since both the principal system and the ancilla are qubits, the optimal initial state of them is given by one of the Bell states, i.e, $\varrho_{SA}=|\Psi\rangle\langle\Psi|$, where $|\Psi\rangle=(|00\rangle+|11\rangle)/\sqrt{2}$. Then, dynamics of the concurrence under the quantum time flip for the depolarizing process is given by
\begin{equation}
C(\varrho_{SA}(t))=\max \left\{0, \frac{4e^{t(1-\omega)}-e^t+1}{3e^t+1} \right\}.
\end{equation}
It is interesting to observe that $C(\varrho_{SA}(t))$ is a monotonically decreasing function of time for all values of the parameter, i.e., $\omega \geq 1/2$. Consequently, entanglement of formation $\mathcal{E}(\varrho_{SA}(t))$, which is a monotonic function of concurrence, exhibits the same behavior displaying no revivals. In Fig.~\ref{fig4}, we show the dynamics of the entanglement of formation for different values of $\omega$, confirming the memoryless nature of the process with respect to the entanglement based measure. Comparison of the trace distance and entanglement based measures of non-Markovianity, both of which depend on the information dynamics between the system and environment, indicates that the kind of memory induced by the quantum time flip for depolarization process can only be captured by the former, never by the latter. We also note that henceforth when we calculate the entanglement based measure, we always take the initial density operator as $\varrho_{SA}=|\Psi\rangle\langle\Psi|$.

\textit{CP-indivisible Process}. The second quantum process that we intend to analyze here is defined by choosing the three real parameters in Eq.~(\ref{pcp}) as 
\begin{equation} \label{cpinc}
\lambda(t)=(1+e^{-\nu t})/2, \quad \lambda_z(t)=e^{- t}, \quad \lambda_*(t)=0.
\end{equation}
This map represents a legitimate unital phase-covariant process if $\nu\geq1$ satisfying the two inequalities in  Eq.~(\ref{cptpc}). The three decoherence rates in the corresponding Lindblad master equation can be obtained using Eq.~(\ref{dr}) as
\begin{equation}
\gamma_{+}(t)=\gamma_{-}(t)=1/2, \quad \gamma_{z}(t)=\frac{1}{4}\left(\frac{2\nu}{e^{\nu t}+1}-1\right).
\end{equation}
Clearly, only the last decoherence rate depends on time, and it is easy to show that $\gamma_{z}(t)$ takes on positive values up until the time $t=\ln(2\nu-1)/\nu$. After this instance, it turns negative and forever remains so. Indeed, if $\nu=1$, this channel reduces to what is known as the eternal non-Markovianity channel in the recent literature~\cite{Hall2014}, since $\gamma_{z}(t)=(-1/4)\tanh[t/2]$, which is negative at all times, implying that the channel is eternally CP-indivisible. 

\begin{figure}
\centering
\includegraphics[width=0.375\textwidth]{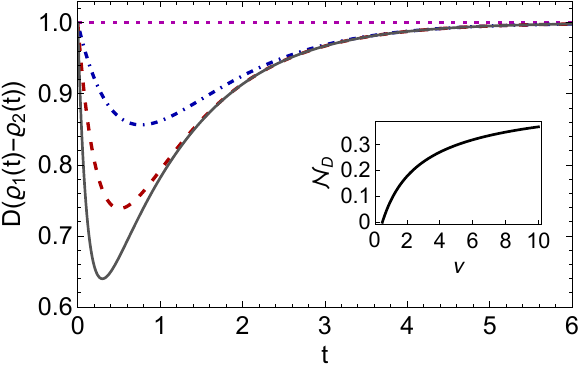}
\caption{Dynamics of the trace distance under the quantum time flip for the CP-indivisible process with the initial state pair $\varrho_1=|+\rangle\langle+|$ and $\varrho_2=|-\rangle\langle-|$, supposing that $\nu=1.0$ (dotted line), $\nu=2.0$ (dot-dashed line), $\nu=4.0$ (dashed line) and $\nu=9.0$ (solid line). The inset shows the degree of memory effects quantified through the trace distance measure $\mathcal{N}_D$ as a function of the channel parameter $\nu$.}
\label{fig5}
\end{figure}

It has been shown that the trace distance based measure witnesses the memory in phase-covariant channels if and only if either one of the conditions 
\begin{equation} \label{tdcon}
\gamma_+(t)+\gamma_{-}(t)+4\gamma_{z}(t) <0, \quad \gamma_+(t)+\gamma_{-}(t) <0,   
\end{equation}
is satisfied~\cite{Teittinen2018}. Obviously, the second inequality cannot be satisfied for the considered channel. In addition, the first inequality implies that $\nu<0$, which is forbidden as the map is CPTP only when $\nu>1$. Hence, the trace distance measure cannot detect the signatures of the memory effects in this process. When it comes to the entanglement based measure, the time-evolution of the concurrence is simply given by $C(\varrho_{SA}(t))=(e^{-t}+e^{-\nu t})/2$, which is a monotonic function of time for all $\nu$. Thus, this CP-indivisible process is viewed as memoryless with respect to both the trace distance and the entanglement based measures of non-Markovianity on its own.

Next, we investigate the consequences of the quantum time flip map for the emergence of memory effects considering the bidirectional CP-indivisible process defined by the three parameters given in Eq.~(\ref{cpinc}). Substituting the Kraus operators $\{M_i\}$ in Eq.~(\ref{qtfk}) with those corresponding to the CP-indivisible phase-covariant channel, results in the quantum time flip map for this dynamical process. Fixing the initial state pair for the principal system as $\varrho_1=|+\rangle\langle+|$ and $\varrho_2=|-\rangle\langle-|$, the dynamics of the trace distance for the considered CP-indivisible process with no definite time direction reads
\begin{equation} \label{tdcpin}
D(\varrho_1(t),\varrho_2(t))=\frac{2e^{t(1-\nu)}+3e^{t}-1}{3e^t+1}.
\end{equation}

\begin{figure}
\centering
\includegraphics[width=0.375\textwidth]{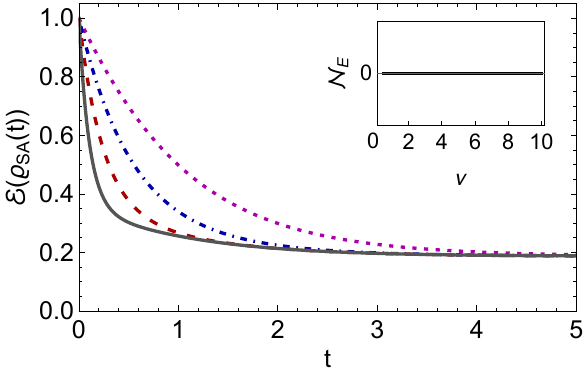}
\caption{Dynamics of the entanglement of formation under the quantum time flip map for the CP-indivisible process for $\nu=1.0$ (dotted line), $\nu=2.0$ (dot-dashed line), $\nu=4.0$ (dashed line) and $\nu=9.0$ (solid line). The inset shows the degree of memory effects quantified by the entanglement based measure $\mathcal{N}_E$ as a function of the channel parameter $\nu$.}
\label{fig6}
\end{figure}

Remarkably, this function behaves monotonically in time only in the case $\nu=1$, which corresponds to the eternal non-Markovianity channel. For $\nu>1$, the trace distance in Eq.~(\ref{tdcpin}) undergoes revivals signalling the presence of memory in the dynamics. Fig.~\ref{fig5} demonstrates the dynamical behavior of $D(\varrho_1(t),\varrho_2(t))$ for different $\nu$ values. At this point, we should mention that, to be able to analytically prove that the quantum time flip map for the CP-indivisible channel is memoryless when $\nu=1$ according to the trace distance measure, one needs to determine the optimal initial state pair in Eq.~(\ref{tdm}). Although we do not have an analytical proof, our numerical simulations strongly suggest that the state pair we consider is indeed optimal. Moreover, we calculate the evolution of the concurrence for the quantum time flip map implemented for the CP-indivisible channel and obtain
\begin{equation} 
C(\varrho_{SA}(t))=\frac{2e^{t(1-\nu)}+e^{t}+1}{3e^t+1},
\end{equation}
which can be easily shown to be a monotonic function of time for all allowed values of the parameter, i.e., $\nu\geq1$. Since this implies a monotonically decaying entanglement of formation as shown in Fig.~\ref{fig6}, we conclude that there exists no memory induced by this quantum time flip map, as quantified by the entanglement based measure.

\subsection{Quantum Switch}

Having discussed the dynamical memory that can be induced by the quantum time flip map, we will now focus on the quantum switch, which implements two quantum processes with no definite causal order. Similarly to the case of the quantum time flip map in the previous part, here we analyze the consequences of the quantum switch for the emergence of memory effects, considering two different families of phase covariant processes. Before starting to present our results, we note that the possibility of observing non-Markovian behaviour, in connection with the violation of the divisibility property, due to quantum switch has been very recently discussed in Refs.~\cite{Maity2023,Anand2023,Mukherjee2023} for two specific instances of unital channels, i.e., isotropic depolarizing and eternal non-Markovianity processes. 

Once again, we set the initial state of the control qubit as the maximally coherent state $\varrho_c=|+\rangle\langle+|$ and measure the control qubit in the coherent basis defined by the pair of projectors $\{|+\rangle\langle+|,|-\rangle\langle-|\}$ to obtain the dynamics of our system of interest. If we assume that the outcome associated with the measurement operator $|+\rangle\langle+|$ occurs, dynamics of the principal system is given by
\begin{equation}
\varrho(t)=\frac{\text{tr}_c[(\mathbb{I}\otimes|+\rangle\langle+|)\mathcal{S}(\varrho\otimes\varrho_c)(\mathbb{I}\otimes|+\rangle\langle+|)]}{\text{tr}
[(\mathbb{I}\otimes|+\rangle\langle+|)\mathcal{S}(\varrho\otimes\varrho_c)]}.
\end{equation}
In the following, we will study the dynamical memory due to quantum switch for non-unital CP-divisible and CP-indivisible phase-covariant maps.

\begin{figure}
\centering
\includegraphics[width=0.375\textwidth]{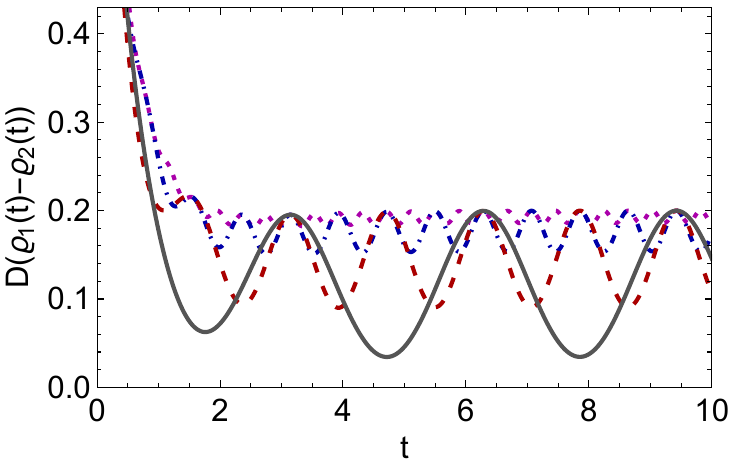}
\caption{Dynamics of the trace distance under the quantum switch for the CP-divisible process with the initial state pair $\varrho_1=|+\rangle\langle+|$ and $\varrho_2=|-\rangle\langle-|$, supposing that $\alpha=8.0$ (dotted line), $\alpha=4.0$ (dot-dashed line), $\alpha=2.0$ (dashed line) and $\alpha=1.0$ (solid line). The trace distance measure $\mathcal{N}_D$ tends to infinity for all positive values of the parameter $\alpha$.}
\label{fig7}
\end{figure}

\textit{CP-divisible Process}. The first map we consider is defined by choosing the parameters in Eq.~(\ref{pcp}) as
\begin{equation} \label{cpinnonu}
\lambda(t)=e^{-t}, \quad \lambda_z(t)=e^{-2 t}, \quad \lambda_*(t)=\frac{2\sin\alpha t}{\sqrt{4+\alpha^2}} ,
\end{equation}
where $\alpha>0$. This quantum channel has been first introduced in Ref.~\cite{Filippov2020} as a curious example of a CP-divisible quantum process, under which the time-evolution of the population terms of the density operator exhibits non-monotonic behavior for any given initial state. The decoherence rates appearing in the Lindblad master equation in Eq.~(\ref{me}) can then be calculated using Eq.~(\ref{dr}) as
\begin{gather}
\gamma_{\pm}(t)=1\pm\frac{(2\sin\alpha t + \alpha \cos \alpha t)}{\sqrt{4+\alpha^2}}, \quad \gamma_z(t)=0.
\end{gather}
Since all three decoherence rates above are non-negative at all times, the process is CP-divisible and hence memoryless with respect to all non-Markovianity quantifiers. In other words, both the trace distance and the entanglement of formation monotonically decay for this process despite the fact that the population terms of the density operator oscillate in time. Besides, as $\gamma_z(t)=0$ throughout the dynamics, this process actually represents a class of generalized amplitude damping channels.

\begin{figure}
\centering
\includegraphics[width=0.375\textwidth]{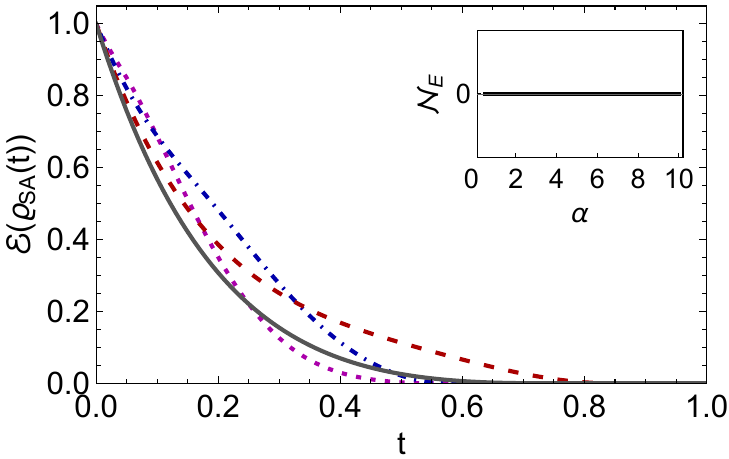}
\caption{Dynamics of the entanglement of formation under the quantum switch for the CP-divisible process for $\alpha=8.0$ (dotted line), $\alpha=4.0$ (dot-dashed line), $\alpha=2.0$ (dashed line) and $\alpha=1.0$ (solid line). The inset shows the degree of memory effects quantified by the entanglement based measure $\mathcal{N}_E$ as a function of the channel parameter $\alpha$.}
\label{fig8}
\end{figure}

Let us explore what happens when the quantum switch is implemented supposing that both quantum channels $\Lambda_1$ and $\Lambda_2$ are chosen identically as the considered generalized amplitude damping channel. That is, substituting the three real parameters given by Eq.~(\ref{cpinnonu}) in the Kraus operators in Eq.~(\ref{switchkraus}) representing the two identical quantum processes $\Lambda_1$ and $\Lambda_2$, we obtain the switched generalized amplitude damping process having no definite causal order. It is quite interesting to note that even though the two quantum processes here are chosen identically, the implementation of quantum switch still gives rise to new dynamics due to the interference between the two channels. Setting the initial state pair of the open system as $\varrho_1=|+\rangle\langle+|$ and $\varrho_2=|-\rangle\langle-|$, we can determine the time-evolution of the trace distance. Although it is straightforward to calculate $D(\varrho_1(t),\varrho_2(t))$ for the switched quantum channel, we omit writing the explicit expression here as it is rather cumbersome. Instead, in Fig.~\ref{fig7}, we show the dynamics of trace distance as a function of time for different values of the channel parameter $\alpha$. Indeed, as also demonstrated in Fig.~\ref{fig7}, $D(\varrho_1(t),\varrho_2(t))$ for this switched process forever oscillates for all $\alpha>0$ between the maximum value of $0.2$ and minimum value of $\alpha^2/(25+4\alpha^2)$. Consequently, we conclude that the quantum switch, when applied to identical maps described by the considered CP-divisible and memoryless generalized amplitude damping channel, results in the emergence of unbounded degree of memory according to the trace distance measure $\mathcal{N}_D$. We also notice that as the parameter $\alpha$ gets larger, the amplitude of oscillations in the trace distance will tend to zero. We stress that, similarly to the case of the quantum time flip, emerging memory in the dynamics here is indeed inserted by the action of the quantum switch superchannel. We next calculate the entanglement of formation considering the same switched process. In Fig.~\ref{fig8}, we display the evolution of $\mathcal{E}(\varrho_{SA}(t))$ plotted for the same set of $\alpha$ values as in Fig.~\ref{fig7}. Since $\mathcal{E}(\varrho_{SA}(t))$ in fact decays monotonically in time for all values of the parameter $\alpha>0$, no dynamical memory could be induced by the quantum switch for this process according to the entanglement based measure. 

\begin{figure}
\centering
\includegraphics[width=0.375\textwidth]{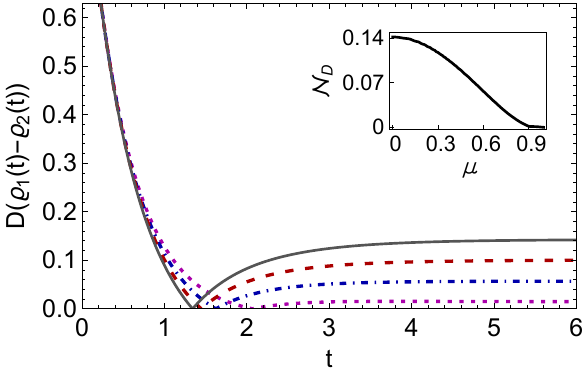}
\caption{Dynamics of the trace distance under the quantum switch for the CP-indivisible process with the initial state pair $\varrho_1=|0\rangle\langle0|$ and $\varrho_2=|1\rangle\langle1|$, supposing that $\mu=0.8$ (dotted line), $\mu=0.6$ (dot-dashed line), $\mu=0.4$ (dashed line) and $\mu=0.0$ (solid line). The inset shows the degree of memory effects quantified through the trace distance measure $\mathcal{N}_D$ as a function of the channel parameter $\mu$.}
\label{fig9}
\end{figure}

\textit{CP-indivisible Process}. The final quantum dynamical map we consider is described by the choice of parameters
\begin{gather} 
\lambda(t)=\frac{1}{2}\sqrt{(1+e^{-t})^2-\mu^2(1-e^{-t})^2 },  \nonumber \\
\lambda_z(t)=e^{-t}, \quad \lambda_*(t)=\mu (1-e^{-t}).\label{cpindivi}
\end{gather}
where $|\mu|<1$ and the requirements for a legitimate quantum dynamics given in Eq.~(\ref{cptpc}) is satisfied. A direct calculation yields that the corresponding decoherence rates in the Lindblad master equation in Eq.~(\ref{me}) are given by
\begin{gather}
\gamma_{\pm}(t)=\frac{1}{2}(1\pm \mu), \quad \gamma_z(t)=\frac{(-1+\mu^2) \sinh t}{4[1+\mu^2+(1-\mu^2)\cosh t]}.
\end{gather}
It can be observed that with the choice of $|\mu|<1$, the two time-independent decoherence rates $\gamma_{\pm}(t)$ are positive. However, the dephasing rate $\gamma_z(t)$ is negative at all times throughout the time evolution, implying an eternally CP-indivisible dynamics~\cite{Filippov2020}. Actually, this quantum channel is a non-unital generalization of the particular case of the unital eternal non-Markovian map we have considered in Eq.~(\ref{cpinc}) with $\nu=1$ since the dephasing rate reduces to $\gamma_{z}(t)=(-1/4)\tanh[t/2]$ for $\mu=0$.

We first demonstrate that although this channel is CP-indivisible for all $t>0$, it is memoryless with respect to the both the trace distance and the entanglement based non-Markovianity measures. Let us recall that the non-monotonic behavior in trace distance dynamics requires that the either one of the inequalities in Eq.~(\ref{tdcon}) is satisfied. As $\gamma_+(t)+\gamma_-(t)=1$, the second inequality clearly cannot be satisfied. Also, the fact that $4\gamma_z\geq-1$ for $|\mu|<1$ at all times implies that the first equality cannot be satisfied either, which proves that the considered channel exhibits no memory according to the trace distance measure. We then calculate the time-evolution of entanglement quantified by concurrence under this non-unital CP-indivisible channel which can be written as
\begin{align} 
C(\varrho_{SA}(t)) =& \frac{1}{2} \left[\sqrt{(e^{-2t}+1)(1-\mu^2)+2e^{-t}(1+\mu^2)} \right. \nonumber \\
                   +& \left. \sqrt{(e^{-t}-1)^2(1-\mu^2)} \right].
\end{align}
For the above function is a monotonically decaying function of time, the process has no memory with respect to the entanglement based measure either. 

\begin{figure}
\centering
\includegraphics[width=0.375\textwidth]{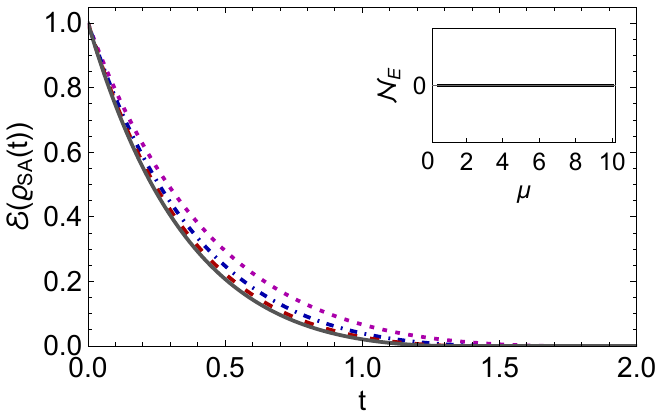}
\caption{Dynamics of the entanglement of formation under the quantum switch for the CP-indivisible process for $\mu=0.8$ (dotted line), $\mu=0.6$ (dot-dashed line), $\mu=0.4$ (dashed line) and $\mu=0.0$ (solid line). The inset shows the degree of memory effects quantified by the entanglement based measure $\mathcal{N}_E$ as a function of the channel parameter $\mu$.}
\label{fig10}
\end{figure}

Recognizing that neither of the memory quantifiers in our study deems the considered CP-indivisible quantum maps as non-Markovian, we explore whether the implementation of the quantum switch for these processes gives rise to dynamical memory. We assume that both $\Lambda_1$ and $\Lambda_2$ are identically set to be the CP-indivisible quantum process defined by the three parameters in Eq.~(\ref{cpindivi}). The switched quantum channel $\mathcal{S}(\varrho \otimes \varrho_c)$ is then given by substituting these three parameters in the Kraus operators in Eq.~(\ref{switchkraus}). To analyze the trace distance measure for the switched channel, we fix the initial state pair of the open system to be $\varrho_1=|0\rangle\langle0|$ and $\varrho_2=|1\rangle\langle1|$, then and calculate the dynamics of the trace distance $D(\varrho_1(t),\varrho_2(t))$. Since the mathematical form of $D(\varrho_1(t),\varrho_2(t))$ turns out to be unwieldy, and therefore not really providing much insight on its own, we choose to merely show its dynamical behavior in Fig.~\ref{fig9} for several allowed values of the parameter $\mu$.
We simply observe that the quantum switch implemented for two identical CP-indivisible channels can induce memory in the dynamics according to the trace distance measure, and the degree of memory measured by $\mathcal{N}_D$ increases as $\mu\rightarrow0$, i.e., as non-unitality of the process weakens. On the other hand, memory in the dynamics tend to vanish while the quantum process resembles more and more to the inverse amplitude damping channel, that is, as $\mu \rightarrow1$. Lastly, in Fig.~\ref{fig10}, we display the dynamical behavior of the entanglement of formation $\mathcal{E}(\varrho_{SA}(t))$ for the switched CP-indivisible channel considering the same values of $\mu$. It is clear that, as in all of the previously analyzed cases of quantum processes with no definite time or causal direction, there emerges no memory here either, due to the implementation of the switch with respect to the entanglement based measure.

\section{Conclusion} \label{sec6}

In summary, we have presented a systematic analysis of dynamical memory effects, quantified by two distinct measures of non-Markovianity, namely the trace distance and the entanglement based measures in quantum processes having no definite time direction and causal order. We have described the open system dynamics of the system using the class of phase-covariant quantum channels, which encompasses processes such as the depolarizing, dephasing and generalized amplitude damping channels.

Quantum processes with indefinite time direction has been employed through the quantum time flip superchannel for certain phase-covariant families of quantum dynamical maps. We have first demonstrated that the CP-invidisible and CP-divisible quantum channels we considered in this part have vanishing memory according to the studied quantifiers of memory effects. Afterwards, we have shown that dynamical memory can emerge in both these types of processes as a consequence of the quantum time flip map, but only according to the trace distance based measure of non-Markovianity. For this measure, we have determined the conditions on the channel parameters that lead to the onset of dynamical memory effects in the time evolution. For instance, we have found out that some degree of anisotropy in the depolarizing channel is necessary for the appearance of memory induced by the indefinite time direction. In addition, for the studied CP-indivisible maps, the only quantum process that fails to have any memory after the application of the quantum time flip is the eternally CP-indivisible channel.

We have implemented the quantum switch superchannel to obtain quantum processes having no definite causal order. In particular, we have chosen the two quantum dynamical maps to be switched in order as identical phase-covariant quantum processes. Similarly to the case of the quantum time flip, we have commenced our investigation by proving that the CP-indivisible and CP-divisible channels in this section are not able to give rise memory effects in the dynamics of the open system, which can be quantified by the two measures of non-Markovianity we studied. Subsequently, we have shown that the implementation of the quantum switch map for the CP-divisible generalized amplitude damping channel induce unbounded degree of memory in dynamics according to the trace distance measure. Furthermore, considering the same non-Markovianity measure, we have also established that the application of the quantum switch for the non-unital generalization of the eternally CP-indivisible maps leads to the generation of memory in the dynamics. It is interesting that no dynamical memory can be generated with respect to the entanglement based measure for the analyzed phase-covariant processes neither by implementing the quantum time flip nor the quantum switch. At this point, we also emphasize that, in the scope of our study, we show that the supermaps that we consider, namely the quantum time flip and the quantum switch, can induce memory in the dynamics of quantum processes. However, this does not necessarily suggest that memory can only be inserted in the open system dynamics through the action of such supermaps, implementing quantum processes having no definite time direction or causal order. In fact, it is potentially possible that less exotic supermaps such as one or two slot causally ordered quantum combs~\cite{Chiribella2008a} can give rise to a similar dynamical behavior regarding memory, as observed in our work.

In closing, we think that our findings can contribute to current understanding of dynamical memory effects and  exotic supermaps such as quantum time flip and quantum switch in a few ways. First, on one hand, memory effects are known to provide some advantages in quantum information processing tasks~\cite{Thorwart2009,Chin2013,Vasile2011,Huelga2012,Chin2012,Bylicka2014,Bylicka2016,Li2020}. On the other hand, the quantum time flip~\cite{Stromberg2022,Guo2022,Liu2023} and quantum switch~\cite{Zhao2020,Chapeau-Blondeau2021,Chiribella2012,Guerin2016,Colnaghi2012,Araujo2014,Renner2022,Ebler2018,Procopio2019,Goswami2020a,Caleffi2020,Bhattacharya2021,Chiribella2021a,Chiribella2021b,Sazim2021,Mukhopadhyay2020,Felce2020,Guha2020,Simonov2022,Liu2022} have been recently shown to offer information theoretic and computational advantages. Hence, the finding that dynamical memory could emerge, when otherwise memoryless quantum channels are superposed via the implementation of such superchannels, is a compelling point which may lead to new research directions. Second, despite the fact that, for instance, the quantum time flip as a supermap cannot be directly implemented, its action on unital channels can be simulated in experiments. Even though the recent experimental works have only implemented unitary gates and measurements in a superposition of the forward and backward time directions~\cite{Stromberg2022,Guo2022}, it should in practice be possible to implement the time flip supermap for unital and memoryless phase-covariant channels and thus experimentally probe the emergence of memory effects in these processes. Lastly, from the perspective of the quantification of memory effects, our results demonstrate curious settings, where the two well-known measures of non-Markovianity, namely, the trace distance and entanglement of formation based measures, seem to consistently disagree.

\begin{acknowledgments}
G. K. and B. \c{C}. are supported by the İzmir University of Economics Research Projects Fund under Grant No. BAP-2023-08. B. \c{C}. and G. K. are also supported by The Scientific and Technological Research Council of Turkey (TUBITAK) under Grant No. 121F246.
\end{acknowledgments}

\bibliography{bibliography}

\end{document}